# Laser synthesis of ruby for photo-conversion of solar spectrum

Aiyyzhy K.O., Barmina E.V., and Shafeev G.A.

Prokhorov General Physics Institute of the Russian Academy of sciences, 38, Vavilov street, 119991, Moscow, Russian Federation

**Abstract**

Ruby grains are synthesized by laser heating of the dry mixture of $Al_2O_3$ and $Cr_2O_3$. Quasi-continuous radiation of a Nd:YAG laser was used for this purpose with average power of 15 W. Synthesized ruby grains demonstrate strong photo-luminescence in the vicinity of 700 nm. Synthesized ruby grains were integrated into a matrix made of flouro-polymer. Luminescene map acquired with the help of a photo-fluorimeter confirms efficient photo-conversion of green-blue radiation of the synthesize ruby into red region. Ruby grans can further be fragmented to smaller particles using the technique of laser fragmentation in liquids.

**Introduction**

Solar radiation is a black body radiation with a maximum at the wavelength of 555 nm. Not all this radiation is used by plants for photosynthesis. The most demanded radiation for efficient proliferation of plants is in red range of spectrum around 700 nm and longer. This is especially true for plants grown in green houses in the terrestrial areas of risky agriculture. Significant efforts were applied for photo-conversion of solar spectrum to red region. Semiconductors based coatings, such as CdS or CdSe could be a possibility to create the coatings that convert solar spectrum to the red region. However, these coatings are not stable under solar light and degrade in a short time. Same concerns the coatings based on organic dyes, they are not photostable. Ruby may be an alternative to semiconductors. Ruby has $Al_2O_3$ lattice in which some $Al^{3+}$ ions are substituted by $Cr^{3+}$ ions. Ruby is highly photo-stable and is characterized by strong photoluminescence in the vicinity of 695 nm [1]. The first laser was realized just on the ruby crystal. This wavelength is of high interest for the use in green houses coatings.

Direct laser synthesis of ruby had already been realized by laser exposure of the mixture of $Al_2O_3$ with $Cr_2O_3$ [2, 3]. The authors used in their work a continuous wave Nd:YAG laser with wavelength of 1064 nm and average power of order of 100 W scanned across the mixture surface. The synthesized ruby crystals were thoroughly characterized by X-ray diffraction technique and Scanning Electron Microscopy and Electron Paramagnetic Resonance techniques.

Infrared characterization of the obtained ruby was also performed. However, visible luminescence of the synthesized ruby crystals was not explored.

Nanoparticles of ruby can be obtained by laser ablation of ruby crystals in liquids [4]. In this case, however, the $Cr^{3+}$ content in these nanoparticles is determined by the composition of initial target material. The most available ruby crystals are bulk laser ruby crystals containing $Cr^{3+}$ at the concentration of 0.05%. For applications as photo-converse coatings the concentration should be higher in order to compensated small size of ruby particles used. Therefore, direct laser synthesis of ruby is preferable way for obtaining ruby particles with desired $Cr^{3+}$ content. The main emphasis of the present work is given to the spectrum of photoluminescence of synthesized ruby particles at various $Cr^{3+}$ content under excitation with various wavelengths.

**Results and discussion**

First, the mixture of industrial $Al_2O_3$ micro-powder with $Cr_2O_3$ micro-powder was prepared with $Cr^{3+}$ content 2, 5, or 8%. The mixture was placed on the bottom of alumina crucible. The thickness of the mixture layer was about 2 mm. The crucible was covered by a soda-lime glass plate. The mixture was exposed to quasi-continuous wave radiation of a Nd:YAG laser with wavelength of 1064 nm in air modulated at frequency of 200 kHz by 1.5 μs pulses with depth of modulation about 30%. The glass cover was far enough distance above the irradiated layer to avoid its cracking due to overheating. Laser beam was scanned across the surface of the layer with the help of F-theta objective and galvo-controlled system of two mirrors. The estimated diameter of the laser spot was 100 μm, scanning velocity of the beam was 100 mm/s. Average power of the laser radiation was 15 W, which corresponds to power density of 150 kW/cm$^2$ on the surface of the mixture. Both components of the mixture are almost transparent at laser wavelength, so the absorption depth is comparable with the thickness of the mixture. The mass of the mixture exposed to laser radiation was around 1 g.

The scanning is accompanied by bright heat emission form the laser-exposed mixture. The mixture was mechanically mixed again after scanning and subjected to another cycle of laser irradiation in the direction perpendicular to the first scanning. The duration of the scanning cycle was about 4 minutes.

Macro-view of ruby grains is shown in Fig. 1. One can see in Fig.1 that agglomerates of ruby are made of spherical particles. This fact suggests that the synthesis of ruby proceeded via melting of the mixture of $Al_2O_3$ and $Cr_2O_3$ and exceeded 2400° C (melting temperature of $Cr_2O_3$).

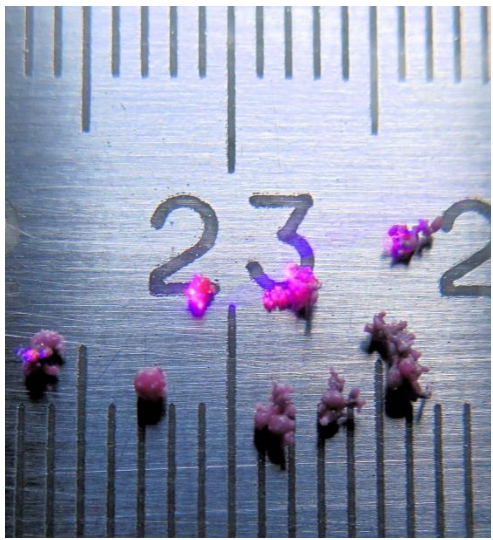

Fig. 1. Macro-view of ruby grains on a millimeter scale. Some grains are illuminated by a laser at wavelength of 405 nm. $Cr^{3+}$ content is of 2%.

Photoluminescence spectra of synthesized ruby grains are shown in Fig. 2.

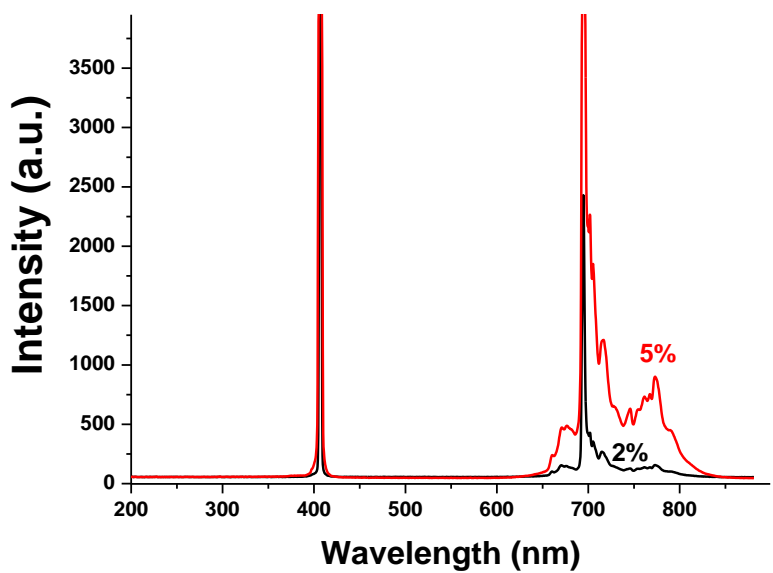

Fig. 2. Photoluminescence (PL) spectra of synthesized ruby under excitation by a laser at wavelength of 405 nm. $Cr^{3+}$ content is 2 (black) and 5% (red). The values of PL intensities are not indicative since the portion of the grains and illumination conditions could be different.

The samples with $Cr^{3+}$ content of 8 -10 % show very weak PL in the same excitation conditions. Apparently, this $Cr^{3+}$ content is the upper limit of efficient PL of a ruby.

The irradiated mixture was then mechanically treated in a mortar to reduce the size of ruby grains. After this some amount of ruby particles (20 mg) was added to fluoro-polymer LF-32 dissolved in acetone. This type of polymers is widely used for green houses coatings. The polymer with ruby loading was evaporated on a glass plate.

The luminescence map of the polymer with ruby was measured with the help of spectro-fluorimeter Jasco. In this flourimeter the excitation of luminescence in a sample can be varied in wide spectral range from UV to infrared. In this work the excitation was varied from 270 to 600 nm. The luminescence map of the fluoro-polymer with synthesized ruby particles is presented in Fig. 3. Also the luminescence map was acquired of both glass substrate with fluoro-polymer and of glass substrate alone. The small peak of luminescence around 690 nm under excitation at 350-370 nm is related to the polymer (label "FP"). The peak at 490 nm under excitation at 450 nm belongs to the glass substrate alone (label "glass"). The two other peaks in red region are related to ruby particles in the polymer.

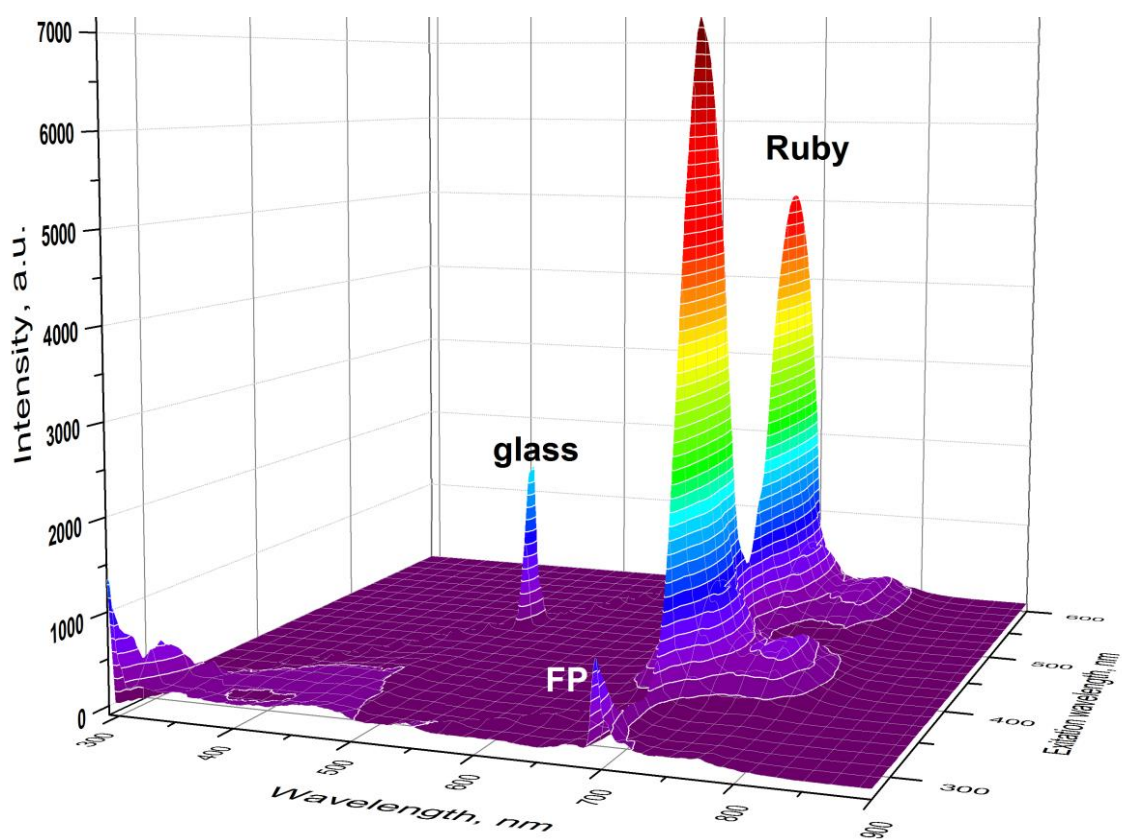

Fig. 3. Luminescence map of ruby particles in fluoro-polymer matrix.

The procedure of laser fragmentation was applied to the synthesized ruby grains to reduce their size. Laser fragmentation is well-established process. It consists in laser exposure of suspension of either micro- or nanoparticles in a liquid [5-8]. The laser beam is focused inside the suspension.

The synthesized ruby grains were subjected to laser fragmentation in isopropanol. In this case the same laser operated in a Q-switch mode, the repetition rate of laser pulses was 10 kHz, pulse duration was 10 nanoseconds and the energy per pulse was 2 mJ. The size of ruby grans is fairly large (see Fig. 1), so the suspension was agitated with the help of rotating glass stylus immersed into suspension. Laser beam was scanned across the entrance window of the cell. The details of experimental setup have been reported elsewhere [7, 8].

Luminescence spectrum of the suspension of fragmented ruby particles in isopropanol is presented in Fig. 4.

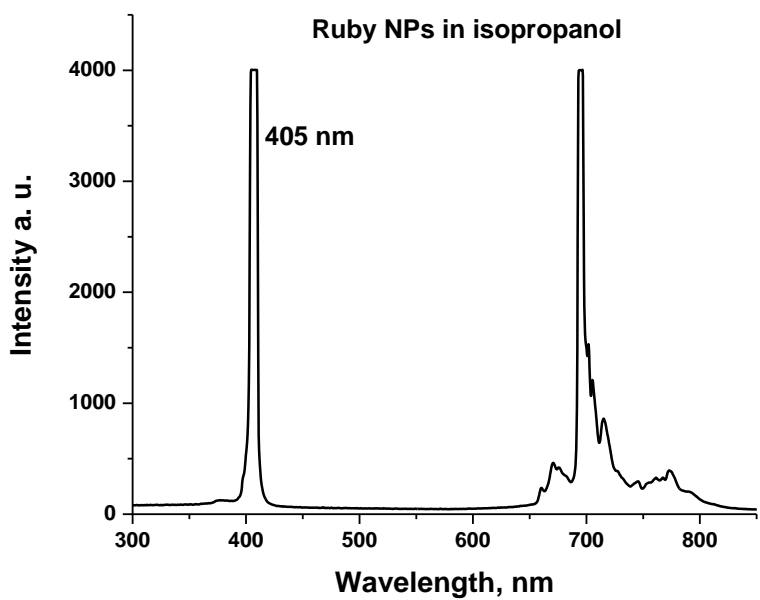

Fig. 4. Photoluminescence spectrum of laser-fragmented ruby grains in isopropanol under excitation by a laser at wavelength of 405 nm. Fragmentation time was of 45 minutes.

The spectrum of laser-fragmented ruby contains the same spectral features as initial dry ruby powder (see Fig. 2).

Thus, the simple and cost-efficient way of obtaining photo-conversion coatings for green houses on the basis of ruby particles has been demonstrated. The synthesized ruby powder shows very efficient conversion of solar radiation in blue-green region of spectrum to required red region of PL. The cost of initial reagents, such as $Al_2O_3$ and $Cr_2O_3$ is low, and the energy of laser consumed during the synthesis is of easily affordable range. The obtained particles are highly photo-stable as stable is ruby in ruby lasers.

**Acknowledgements**

This work was supported by a grant of the Ministry of Science and Higher Education of the Russian Federation (075-15-2022-315) for the organization and development of a World-class research center "Photonics". The authors thank I.V. Baymler for his help in data curation.